# Reconfigurable all-optical inference via tunable second-harmonic generation and spin-orbit coupling cascade

Li Zhang,[1] Zikuan Zhuang,[1] Ronghao Deng,[2] Ling Hong,[1] Yu Zhang,[1] Fei Lin,[1] Zhengxian Liu,[1] Jingxuan Sun,[1] Wenguo Zhu[2]*, Zhenwei Xie[3]*, Yongyao Li,[1] Dongxu Zhao,[1] and Xiaocong Yuan[3]*

[1]Guangdong-HongKong-Macao Joint Laboratory for Intelligent Micro-Nano Optoelectronic Technology, School of Physics and Optoelectronic Engineering, Foshan University, Foshan 528225, China
[2]Key Laboratory of Optoelectronic Information and Sensing Technologies of Guangdong Higher Education Institutes, Department of Optoelectronic Engineering, Jinan University, Guangzhou, 510632, China
[3]Nanophotonics Research Center, Shenzhen Key Laboratory of Micro-Scale Optical Information Technology, Institute of Microscale Optoelectronics &State Key Laboratory of Radio Frequency Heterogeneous Integration, Shenzhen University, Shenzhen, China
*Corresponding Author: zhuwg88@163.com; ayst31415926@szu.edu.cn; xcyuan@szu.edu.cn

**Abstract**
Spin-orbit coupling (SOC) is widely exploited as a fundamental mechanism for generating orbital angular momentum (OAM); however, conventional approaches typically lack flexibility and tunability. Here, we introduce a continuously tunable second-harmonic generation (SHG)-SOC cascade mechanism modulated by a spatially movable nonlinear crystal. Under linearly polarized excitation, the SHG-SOC cascade engages synchronously with both degenerate and nondegenerate SHG processes, thereby expanding the OAM spectrum and significantly enhancing the information density and feature-mapping capacity of the optical field. Moreover, the OAM spectral distribution can be continuously reconfigured simply by translating the nonlinear crystal. This deterministic physical evolution, which maps simple OAM modes onto a tunable high-dimensional OAM space, is mathematically analogous to the high-dimensional feature expansion performed by a kernel function of a support vector machine (SVM) in machine learning. Such dimensional expansion can project linearly inseparable input data into a high-dimensional space where they become linearly separable. Exploiting this physics-algorithm analogy, we develop a reconfigurable all-optical inference platform. As a proof of concept, we successfully perform classification tasks, including the recognition of Iris flowers and Palmer penguins. This work establishes a scalable, physically reconfigurable architecture for high-dimensional all-optical computing and neuromorphic photonics.

## INTRODUCTION

The spin angular momentum (SAM) and orbital angular momentum (OAM) of photons, corresponding to the polarization state and the helical phase of light respectively, constitute two fundamental degrees of freedom for tailoring light-matter interactions (*1*, *2*). The interplay between these two degrees of freedom, i.e., optical spin-orbit coupling (SOC), not only reveals the microscale dynamics of optical fields but has also emerged as a core mechanism for constructing complex structured light (*3*-*6*). Linear SOC platforms based on tightly focused beams (*7*, *8*), anisotropic media (*9*-*11*) and metasurfaces (*12*-*14*) have driven significant progress across diverse fields, including high-precision metrology (*15*, *16*), optical communications (*17*, *18*), microscopy (*19*, *20*), sensing (*21*, *22*), and optical manipulation (*23*, 2*4*). However, such schemes rely on rigid architectures and predetermined SOC channels, yielding finite-dimensional spin-OAM superpositions and precluding dynamic mode reconfigurability.

Recently, integrating SOC with nonlinear optical processes, such as second-harmonic generation (SHG), parametric down-conversion and high-harmonic generation, has provided new physical pathways for the frequency conversion and dimensional expansion of structured light (*25-30*). For instance, in the nonlinear-crystal-based SHG-SOC cascade, the spin-OAM modes of the incident light can be mapped into new OAM modes of second-harmonic photons, synergistically governed by nonlinear optical selection rules and crystal symmetries (*31-34*). In contrast to linear counterparts, such SHG-SOC cascade mechanisms introduce additional controllable degrees of freedom, including phase-matching conditions, nonlinear interaction lengths, and spatial beam profiles, thereby promising highly flexible manipulation of OAM. In practice, however, current schemes rely almost entirely on purely circularly polarized inputs, which confines the polarization evolution to singular spin paths, inevitably yielding a sparse and predominantly discrete OAM spectrum. Moreover, the inherently rigid design of the current SHG-SOC cascade fundamentally lacks dynamic reconfigurability. Such single-state excitations preclude the generation of a multitude of orthogonal modes, leaving the vast potential of harnessing nonlinear optics to construct high-dimensional feature spaces largely untapped. Therefore, it remains a challenge to generate a continuously tunable high-dimensional OAM spectrum using simple polarization inputs.

Here, we demonstrate a continuously tunable cascaded SOC-SHG schemes within a single, spatially movable nonlinear crystal. Under spatially uniform or nonuniform linearly polarized excitation propagating along the optical axis of the crystal, SHG and SOC interact with each other. This gives rise to an SHG process that includes not only the self-coupled degenerate SHG of the fundamental frequency (FF) field and its corresponding SOC counterpart, but also the cross-coupled nondegenerate SHG arising from their mutual interaction. As the beam propagates longitudinally through the medium, this synergistic nonlinear evolution broadly expands the OAM spectrum into a high-dimensional state space. This broad multimode output not only dramatically enhances the information density of the localized optical field but also provides a vast and complete orthogonal basis for feature mapping. Crucially, such a scheme offers a high degree of manipulation freedom: simply translating the nonlinear crystal along the optical axis enables the continuous, precise reconfiguration of the mode amplitude distribution within this superposition state.

Furthermore, such physical evolution maps low-order linearly polarized input states onto highly tunable, high-dimensional OAM projections, which is mathematically isomorphic to high-dimensional feature expansion of support vector machine (SVM) in machine learning (*35-39*). In classical SVM algorithms, this crucial dimensionality expansion is performed by a specific mathematical kernel function. Here, as shown in Fig. 1A, we exploit the SHG-SOC process to physically implement an optical analog of the kernel function. This scheme projects linearly inseparable input data into a high-dimensional space where they become linearly separable. Leveraging this physics-algorithm analogy, we construct a reconfigurable all-optical inference platform. Within this architecture, the evolution of the cascaded SHG-SOC performs high-dimensional feature extraction, while the longitudinal position of the nonlinear crystal serves as a tunable parameter that directly modulates the optical kernel function. As a proof of concept, we experimentally execute classification tasks, successfully classifying the Iris flower and Palmer penguins datasets. This work not only deepens the understanding of spin-orbit interactions in nonlinear regimes but also introduces a hardware-efficient, low-power photonic platform, paving the way for all-optical inference computing based on intrinsic physical evolution.

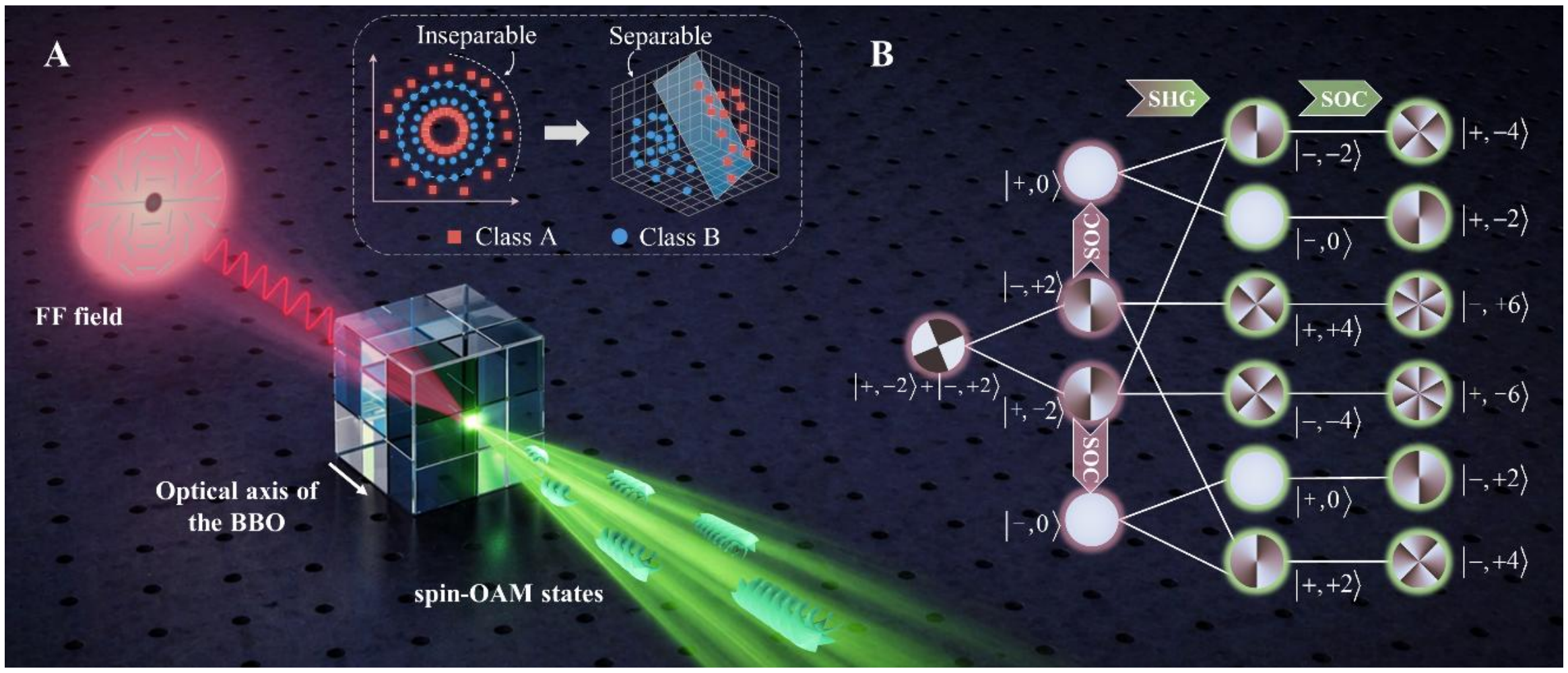

**Fig. 1. Schematic of reconfigurable all-optical inference via continuously tunable SHG-SOC cascade.** (**A**) Conceptual graphs of reconfigurable all-optical inference via continuously tunable SHG-SOC cascade in a nonlinear crystal; (**B**) Schematic diagram of the spin-OAM states evolution via SHG-SOC cascade.

# RESULTS

## The evolution of SHG-SOC cascade

Assume an FF paraxial field propagates in a nonlinear crystal. The FF field will excite second-harmonic frequency (SF) fields. Each frequency field in the crystal should obey the following equation (*40*):

$$\nabla^2 \mathbf{E}_n - \nabla(\nabla \cdot \mathbf{E}_n) - \frac{\omega_n^2}{c^2} \overset{\leftrightarrow}{\varepsilon}(\omega_n) \mathbf{E}_n = -\frac{\omega_n^2}{\varepsilon_0 c^2} \mathbf{P}_n^{NL} , \tag{1}$$

where the relative permittivity of a uniaxial crystal is in the form of $\overset{\leftrightarrow}{\varepsilon}(\omega_n) = dig\left[n_o^2(\omega_n),\, n_o^2(\omega_n),\, n_e^2(\omega_n)\right]$ with $\omega_n$ being the frequency of field. The slowly varying amplitude can be extracted from the light field $\boldsymbol{E}_n = \exp\left[ik_n n_o(\omega_n) z\right]\boldsymbol{A}_n$, where $\boldsymbol{A}_n = \boldsymbol{A}_o^n + \boldsymbol{A}_e^n$. $\boldsymbol{A}_{o,e}^n$ are the ordinary and extraordinary wave components that propagate independently in a uniaxial crystal. The nonlinear polarization is $\boldsymbol{P}_{\omega_1+\omega_2}^{NL} = \overset{\leftrightarrow}{d} \cdot \boldsymbol{E}^{\omega_1} \boldsymbol{E}^{\omega_2}$. For a 3m point group nonlinear crystal, the nonlinear optical tensor is in the form of (*41*)

$$\overset{\leftrightarrow}{d} = \begin{bmatrix} d_{11} & -d_{11} & 0 & 0 & d_{15} & 0 \\ 0 & 0 & 0 & d_{15} & 0 & -d_1 \\ d_{13} & d_{13} & d_{33} & 0 & 0 & 0 \end{bmatrix}, \tag{2}$$

For paraxial beams, the $z$-component of the light fields is negligibly small. The $z$-component of $P_{\omega_1+\omega_2}^{NL}$ and the terms of $P_{\omega_1+\omega_2}^{NL}$ containing the $z$-component of the electric fields can be neglected. In the circular basis, Eq (1) can be rewritten as (detailed Section I and Section II in the Supplementary Materials):

$$2i\frac{\partial A_+^{\omega}}{\partial z} + \frac{\chi(\omega)}{2k_1 n_o(\omega)}\left[\frac{\partial^2}{\partial x^2} + \frac{\partial^2}{\partial y^2}\right]A_+^{\omega} = -\frac{\delta}{2k_1 n_o(\omega)}\left[\frac{\partial}{\partial x} - i\frac{\partial}{\partial y}\right]^2 A_-^{\omega} , \tag{3a}$$

$$2i\frac{\partial A_-^{\omega}}{\partial z} + \frac{\chi(\omega)}{2k_1 n_o(\omega)}\left[\frac{\partial^2}{\partial x^2} + \frac{\partial^2}{\partial y^2}\right]A_-^{\omega} = -\frac{\delta}{2k_1 n_o(\omega)}\left[\frac{\partial}{\partial x} + i\frac{\partial}{\partial y}\right]^2 A_+^{\omega} , \tag{3b}$$

$$2i\frac{\partial A_+^{2\omega}}{\partial z}+\frac{\chi(2\omega)}{2k_2n_o(2\omega)}[\frac{\partial^2}{\partial x^2}+\frac{\partial^2}{\partial y^2}]A_+^{2\omega}=-\frac{\delta}{2k_2n_o(2\omega)}[\frac{\partial}{\partial x}-i\frac{\partial}{\partial y}]^2A_-^{2\omega}-\frac{2\sqrt{2}d_{11}k_2}{n_o(2\omega)}(A_-^{\omega})^2\exp\{ik_2[n_o(\omega)-n_o(2\omega)]z\},\tag{3c}$$

$$2i\frac{\partial A_-^{2\omega}}{\partial z}+\frac{\chi(2\omega)}{2k_2n_o(2\omega)}[\frac{\partial^2}{\partial x^2}+\frac{\partial^2}{\partial y^2}]A_-^{2\omega}=-\frac{\delta}{2k_2n_o(2\omega)}[\frac{\partial}{\partial x}+i\frac{\partial}{\partial y}]^2A_+^{2\omega}-\frac{2\sqrt{2}d_{11}k_2}{n_o(2\omega)}(A_+^{\omega})^2\exp\{ik_2[n_o(\omega)-n_o(2\omega)]z\}.\tag{3d}$$

$A_\pm^{\omega,2\omega}$ are the right- and lelf-handed circular polarzation (RCP and LCP) components of slowly varying amplitudes for FF and SF fields. $\chi(\omega,2\omega)=n_o^2(\omega,2\omega)/n_e^2(\omega,2\omega)+1$, $k_{1,2}$ are the wavenumbers of FF and SF waves in the vacuum. $d_{11}$ is the second-harmonic coefficient of the nonlinear crystal. In the small signal approximation, the propagation of the FF field is not affected by the nonlinear process. The dymanics of the FF field can be obtained by solving Eqs. (3a) and (3b) as the input field is given. In the polar coordinates, the RCP and LCP components of the FF field are (*42*)

$$A_+(r,\phi,z)=\sum_n\{\exp(in\phi)F_+^n(r,z)+\exp[i(n'-2)\phi]G_-^{n'-2}(r,z)\},\tag{4a}$$

$$A_-(r,\phi,z)=\sum_n\{\exp(in'\phi)F_-^{n'}(r,z)+\exp[i(n+2)\phi]G_+^{n+2}(r,z)\},\tag{4b}$$

where

$$F_\pm^n(r,z)=\pi\int dkk\left[\exp(-ik_\perp^2z/2k_0n_o)+\exp(-in_ok_\perp^2z/2k_0n_e^2)\right]J_n(kr)\tilde{E}_\pm^n(k),$$

$$G_\pm^n(r,z)=\pi\int dkk\left[\exp(-ik_\perp^2z/2k_0n_o)-\exp(-in_ok_\perp^2z/2k_0n_e^2)\right]J_n(kr)\tilde{E}_\pm^{n\mp2}(k).$$

$\tilde{E}_+^n(k)$ and $\tilde{E}_-^{n'}(k)$, which are detailed in the Supplementary Materials, represent the angular spectrum of incident field $E_\pm(r,\phi,0)$. One can find from Eqs. (4a) and (4b) that, the terms containing $F_\pm^n(r,z)$ carry the same optical vortex as the incident field. The terms with $G_\pm^n(r,z)$ describe the SOC effect in the uniaxial crystal, where the optical vortex charges increase or decrease 2-order.

According to Eqs. (3c) and (3d), the RCP and LCP components of the FF field serve as the sources of the LCP and RCP SF fields. It is difficult to solve Eqs. (3c) and (3d) directly. The source can be considered as the incident SF fields at the $z$ plane, thus Eqs. (3c) and (3d) can be rewritten as

$$2i\frac{\partial A_+^{2\omega}(z>z_0)}{\partial z}+\frac{\chi}{2k_2n_o(2\omega)}[\frac{\partial^2}{\partial x^2}+\frac{\partial^2}{\partial y^2}]A_+^{2\omega}(z>z_0)=-\frac{\delta}{2k_2n_o(2\omega)}[\frac{\partial}{\partial x}-i\frac{\partial}{\partial y}]^2A_-^{2\omega}(z>z_0),\tag{5a}$$

$$2i\frac{\partial A_-^{2\omega}(z>z_0)}{\partial z}+\frac{\chi}{2k_2n_o(2\omega)}[\frac{\partial^2}{\partial x^2}+\frac{\partial^2}{\partial y^2}]A_-^{2\omega}(z>z_0)=-\frac{\delta}{2k_2n_o(2\omega)}[\frac{\partial}{\partial x}+i\frac{\partial}{\partial y}]^2A_+^{2\omega}(z>z_0),\tag{5b}$$

with the boundary conditions of

$$A_+^{2\omega}(z=z_0)=-\frac{2\sqrt{2}d_{11}k_2}{n_o(2\omega)}(A_-^{\omega}(x,y,z=z_0)^2\exp\{ik_2[n_o(\omega)-n_o(2\omega)]z_0\},$$

$$A_-^{2\omega}(z=z_0)=-\frac{2\sqrt{2}d_{11}k_2}{n_o(2\omega)}(A_+^{\omega}(x,y,z=z_0)^2\exp\{ik_2[n_o(\omega)-n_o(2\omega)]z_0\}.$$

The SF field in the output plane should be obtained by integrating all the SF field excited at different $z$ planes.

$$A_{+}^{2\omega}(z=d)=\int_0^d A_{+}^{2\omega}(z>z_0)dz_0\ , \tag{6a}$$

$$A_{-}^{2\omega}(z=d)=\int_0^d A_{-}^{2\omega}(z>z_0)dz_0\ . \tag{6b}$$

By solving Eqs. (5a) and (5b) under the boundary conditions, we obtain

$$\begin{aligned}A_{+}^{2\omega}(r,\phi,z>z_0)=\sum_{n,n'}\{&\exp[i2n'\phi]F_{2\omega,+}^{n',n'}+\exp[i(2n+4)\phi]F_{2\omega,+}^{n+2,n+2}+\exp[i(n+n'+2)\phi]F_{2\omega,+}^{n',n+2}\\&+\exp[i(2n-2)\phi]G_{2\omega,-}^{n,n}+\exp[i(2n'-6)\phi]G_{2\omega,-}^{n'-2,n'-2}+\exp[i(n+n'-4)\phi]G_{2\omega,-}^{n,n'-2}\}\end{aligned}\ , \tag{7a}$$

$$\begin{aligned}A_{-}^{2\omega}(r,\phi,z>z_0)=\sum_{n,n'}\{&\exp[i2n\phi]F_{2\omega,-}^{n,n}+\exp[i(2n'-4)\phi]F_{2\omega,-}^{n'-2,n'-2}+\exp[i(n+n'-2)\phi]F_{2\omega,-}^{n,n'-2}\\&+\exp[i(2n'+2)\phi]G_{2\omega,+}^{n',n'}+\exp[i(2n+6)\phi]G_{2\omega,+}^{n+2,n+2}+\exp[i(n+n'+4)\phi]G_{2\omega,+}^{n',n+2}\}\end{aligned}\ , \tag{7b}$$

where

$$F_{2\omega,\pm}^{n,n'}(r,z)=\pi\int dkk\left[\exp(-ik_\perp^2z/2k_0n_o)+\exp(-in_ok_\perp^2z/2k_0n_e^2)\right]J_{n+n'}(kr)\tilde{E}_{2\omega,\pm}^{n,n'}(k)\ ,$$

$$G_{2\omega,\pm}^{n,n'}(r,z)=\pi\int dkk\left[\exp(-ik_\perp^2z/2k_0n_o)-\exp(-in_ok_\perp^2z/2k_0n_e^2)\right]J_{n+n'\pm2}(kr)\tilde{E}_{2\omega,\pm}^{n,n'}(k)\ .$$

$\tilde{E}_{2\omega,\pm}^{n,n'}(k)$ is the angular spectrum of the excited SF field at the $z=z_0$ plane, which is detailed in the Supplementary Materials. $\tilde{E}_{2\omega,\pm}^{n,n'}(k)$ contains the SOC effect in the FF field, while the SOC effect in the SF field is described by $G_{2\omega,\pm}^{n,n'}$. The first terms in Eqs. (7a) and (7b), carrying $2n'$ or $2n$ order optical vortex, are directly degenerate SHG of incident field without SOC processes in FF fields. The second terms in Eqs. (7a) and (7b), carrying $2n+4$ or $2n'-4$ order optical vortex, are the degenerate SHG field of spin-orbit converted FF fields. The third terms in Eqs. (7a) and (7b), carrying $n+n'\pm2$ order optical vortex, are the nondegenerate SHG field of a spin-orbit converted and a spin-orbit unconverted FF fields. Additional SOC processes occur in the SHG field for the last three terms of Eqs. (7a) and (7b). Therefore, different second harmonic spin-OAM states can be created, as shown in Fig. 1B.

When a FF field is focused into the nonlinear crystal (see Fig. 2 (A to C)), the SF field excited in the focal plane is strongest and makes much contribution to the output SF field. Thus, we can tune the spin-OAM states of the SF field by varying the focal position of the FF field $d_F$. Assume the incident FF field is a High-Order Poincaré (HOP) beam $\mathbf{E}_{in}=|+,-2\rangle+|-,+2\rangle$ with $|\pm\rangle$ and $|\pm2\rangle$ being respectively the SAM states and the OAM states (*43, 44*). The intensity distributions of FF and SF fields in the input, middle, and output planes are given in Fig. 2 (A to C) for the focal positions $d_F$=0, $d$/2, and $d$, respectively. Both the FF and SF fields emerge as the smallest spots at the focal positions.

The SOC effect occurs in the FF fields, which however is independent of the focal position. This is because the SOC process relies on the angular spectrum of the incident fields (see Eqs. (4a) and (4b)). The SOC process induces $|-,0\rangle$ and $|+,0\rangle$ states. As the FF field propagates in the crystal, the energy of $|\pm,\mp2\rangle$ states converts gradually to the $|\mp,0\rangle$ states, while the total energy of FF field is conserved (see Fig. 2 (D to F)).

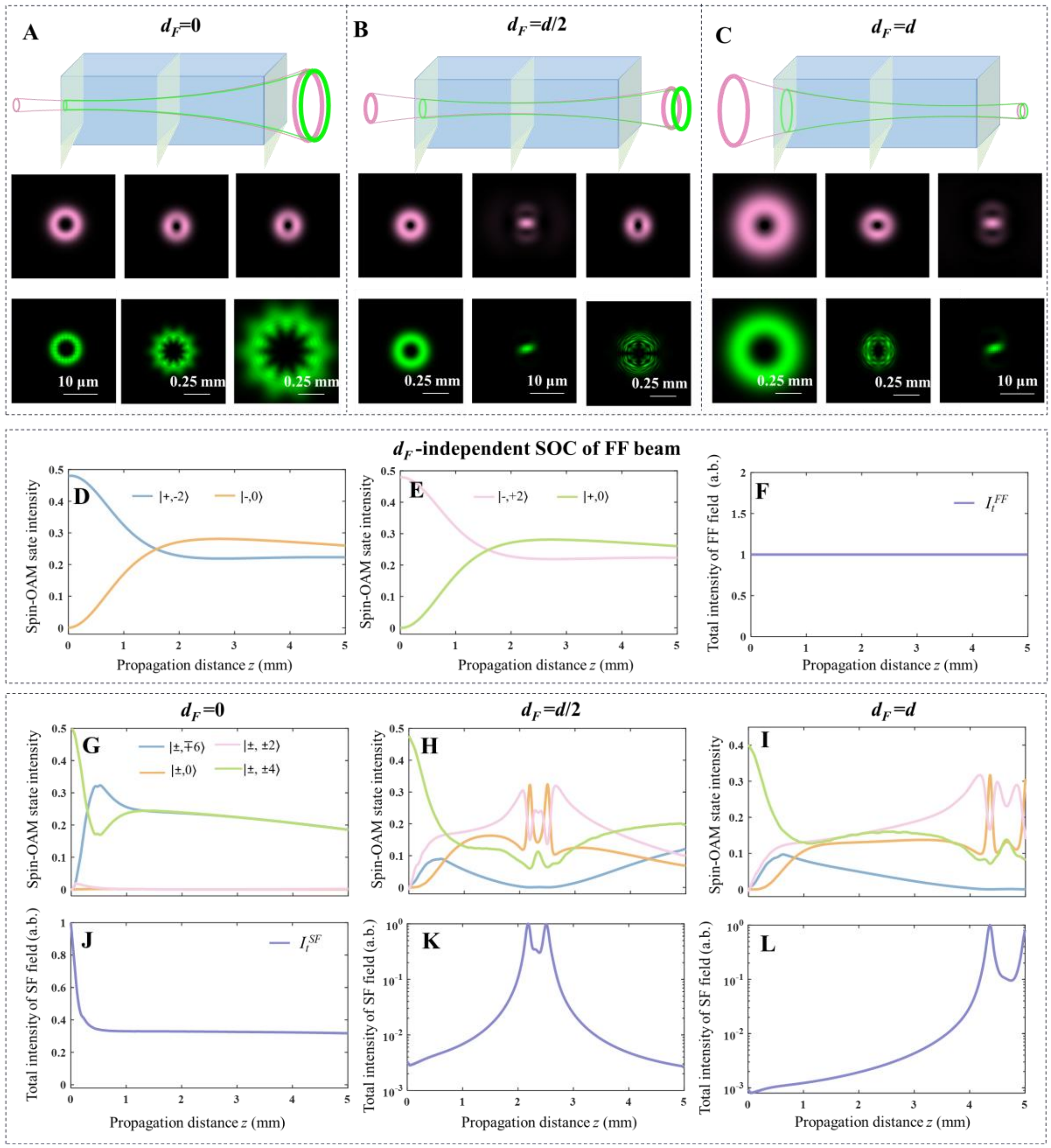

**Fig. 2. The evolution results of SHG-SOC cascade in a nonlinear crystal.** (**A** to **C**) Schematics of SF field generations for the focal positions of FF field $d_F$ in the input (A), middle (B), and output (C) planes of the crystal, where the corresponding intensity distributions of FF and FF fields are given. (**D** to **F**) Intensities of spin-OAM states (D, E) and total intensity (F) of FF fields varying with the propagation distance. (**G** to **L**) Intensities of spin-OAM states (G to I) and total intensities (J to L) of SF fields as functions of propagation distance for $d_F$=0 (G and J), *d*/2 (H and K), and *d* (I and L), respectively.

According to Eqs. (7a) and (7b), there are 12 spin-OAM states in the SF field when the incident FF field is $\mathbf{E}_{in} = |+,-2\rangle + |-,+2\rangle$. These states include $|\pm,\pm4\rangle$, $|\pm,0\rangle$, $|\pm,\pm2\rangle$, $|\pm,\mp6\rangle$, $|\pm,\mp2\rangle$, and $|\pm,\mp4\rangle$, which undergo different processes. States $|\pm,\pm4\rangle$ are generated directly via SHG from the incident FF field, while states $|\pm,0\rangle$ go through the SOC-SHG process. States $|\pm,\pm2\rangle$ and $|\pm,\mp6\rangle$ go through the SOC-SHG and SHG-SOC processes, respectively. States $|\pm,\mp2\rangle$ and $|\pm,\mp4\rangle$ go through the SOC-SHG-SOC and SOC-SHG-SOC processes, respectively. In our case, the spin-OAM states undergoing three processes, i.e. $|\pm,\mp2\rangle$ and $|\pm,\mp4\rangle$, possess very small intensity, and thus are unplotted in Fig. 2 (G to I).

When the focal position of the FF beam is at the input plane of the crystal ($d_F$=0), the $|\pm,\pm4\rangle$ states are generated in the SF field, which originates directly from the SHG of incident FF field. As shown in Fig. 2G, the $|\pm,\pm4\rangle$ states convert gradually into the $|\pm,\mp6\rangle$ states via the SOC process in the SF field. The intenisties of $|\pm,\pm4\rangle$ and $|\pm,\mp6\rangle$ states are identical for the propagation distance $z$>1.2 mm, while the total intensity of the SF field remains constant when $z$>0.5 mm (see Fig. 2J). When the FF beam is focused at the crystal's midplane ($d_F$=$d$/2), the total SF field intensity peaks near this plane, while dropping by two orders of magnitude at the output plane (see Fig. 2K). As shown in Fig. 2H, the spin-OAM intensities exhibit dynamic variations near the focal plane. The output SF field contains nonzero $|\pm,\pm4\rangle$, $|\pm,0\rangle$, $|\pm,\pm2\rangle$, and $|\pm,\mp6\rangle$ states. When the focal plane is moved to the output plane of the crystal ($d_F$=$d$), both the intensities of spin-OAM states and the total intensity undergo dynamic variations near the output plane, as shown by Fig. 2 (I and L). The output SF field contains nonzero $|\pm,\pm4\rangle$, $|\pm,0\rangle$, and $|\pm,\pm2\rangle$ states, with the $|\pm,\mp6\rangle$ states vanishing. Thus, the spectra of spin-OAM states of the output SF field can be flexibly tuned by controlling the focal position of the FF field.

**Spin-OAM spectrum of SF field**

The focal-position-dependent spin-OAM spectrum of the SF fields generated by a HOP ($\mathbf{E}_{in}=|+,-2\rangle+|-,+2\rangle$) for FF beam, is investigated experimentally (for the experimental setup, see the MATERIALS AND METHODS). Fig. 3 (A to C) demonstrates the measured results of normalized spin-OAM spectra at three representative intra-crystal focal positions $d_F$. With the focus positioned at the front surface of the crystal ($d_F$=0.000 mm), the output spectrum is dominated by the $|\pm,\pm4\rangle$ and $|\pm,\mp6\rangle$ states. This indicates that the primary contributing mechanisms are SHG and SOC-SHG. As the focus position is advanced to the inside of the crystal ($d_F$=3.125 mm), the SF spin-OAM spectrum becomes more complex, with all even-integer OAM modes from $n$=0 to $n$=6 being prominently detected. This signifies a complex interplay of concurrent processes, including degenerate SHG, nondegenerate SHG and SHG-SOC. When the focus is located at the exit face ($d_F$=5.000 mm), the $|\pm,0\rangle$ states become the dominant component, while the $|\pm,\pm2\rangle$ and $|\pm,\pm4\rangle$ states persist with minor contributions. This outcome is primarily attributed to a different cascaded process where SOC precedes the SHG.

To provide a more comprehensive picture, the continuous evolution of each spin-OAM mode along with the focal position is quantitatively analyzed. As shown in Fig. 3D, the evolutionary trajectories of the spin-OAM states are distinct and highly dependent on the focus location. The powers of the $|\pm,\pm4\rangle$ and $|\pm,\mp6\rangle$ states remain high and relatively stable throughout the first half of the crystal before gradually attenuating in the second half. In contrast, the power of $|\pm,\pm2\rangle$ component steadily grows through the first half, reaching a plateau before decaying rapidly as the focus nears the exit face. The $|\pm,0\rangle$ component exhibits the most complex evolution: its power first increases; then decreases; and finally experiences a sharp rise just as the focus approaches the exit of the crystal. The strong agreement between the experimental data and the simulation not only corroborates our physical model but also validates that precise control over the intra-crystal focal position is a viable and effective approach for tuning the output spin-OAM spectrum distribution.

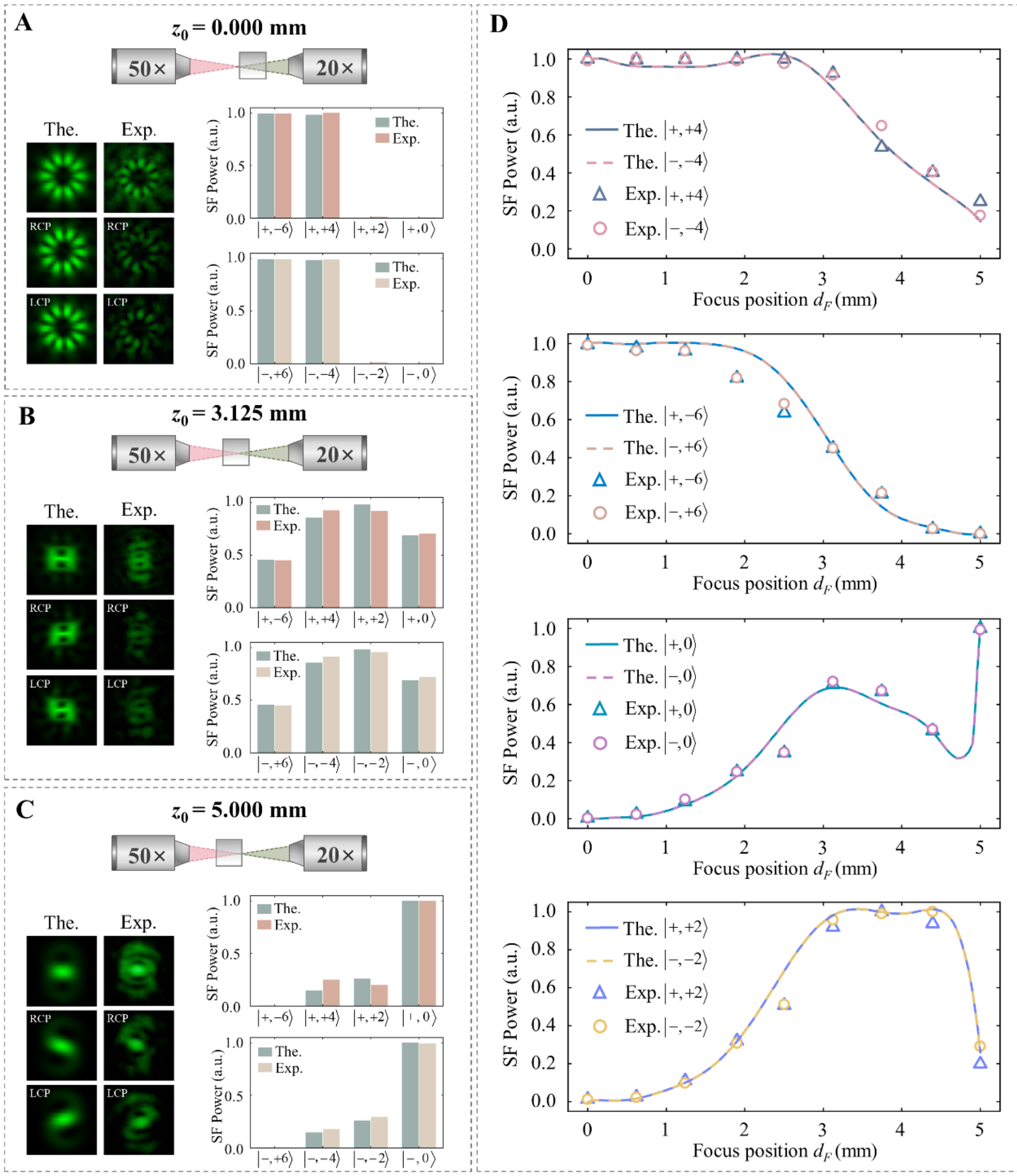

**Fig. 3. Spin-OAM states of SF field for HOP incident FF beams.** (**A** to **C**) Experimental setup as well as intensity patterns of SF fields and spin-OAM spectra for the focus positions at $d_F$ =0.000 mm, 3.125 mm, and 5.000 mm, respectively; (**D**) Theoretical and experimental SF power evolution along with focus position $d_F$.

Furthermore, the dependence of the spin-OAM spectrum of the SF field on the longitudinal focal position $d_F$ is investigated for a linearly polarized FF beam. Fig. 4A elucidates the continuous, $d_F$-dependent evolution of the spin-OAM spectrum. The excellent agreement between theoretical modeling and experimental measurements confirms that the generation efficiency for distinct OAM components is highly $d_F$-dependent. Specifically, the $|-,+4\rangle$, $|+,+4\rangle$, $|-,+6\rangle$, $|+,+2\rangle$ states exhibit maximum efficiency near $d_F$ =0.000 mm and gradually attenuate as $d_F$ increases. Conversely, the $|+,+6\rangle$ and $|-,+2\rangle$ states are suppressed at $d_F$=0.000 mm but grow through the first half of the medium, reaching a plateau before decaying as the focus approaches the exit face. Interestingly, the $|-,0\rangle$ and $|+,+8\rangle$ states exhibit more complex dynamics: the $|+,+8\rangle$ power increases and then decreases, while the $|-,0\rangle$ power increases, momentarily decreases, and subsequently experiences a sharp rise.

Fig.4B demonstrates a state-resolved decomposition of the SHG signal at three representative focal positions ($d_F$ =0.000 mm, 3.125 mm, and 5.000 mm). This visualization, which resolves the SHG power by both OAM and SAM, confirms the $d_F$-dependent spectral reshaping observed in Fig. 4A. These results collectively establish that by translating the nonlinear medium along the propagation axis of a focused fundamental beam, one can deterministically engineer the spin-OAM spectrum of the generated SF field.

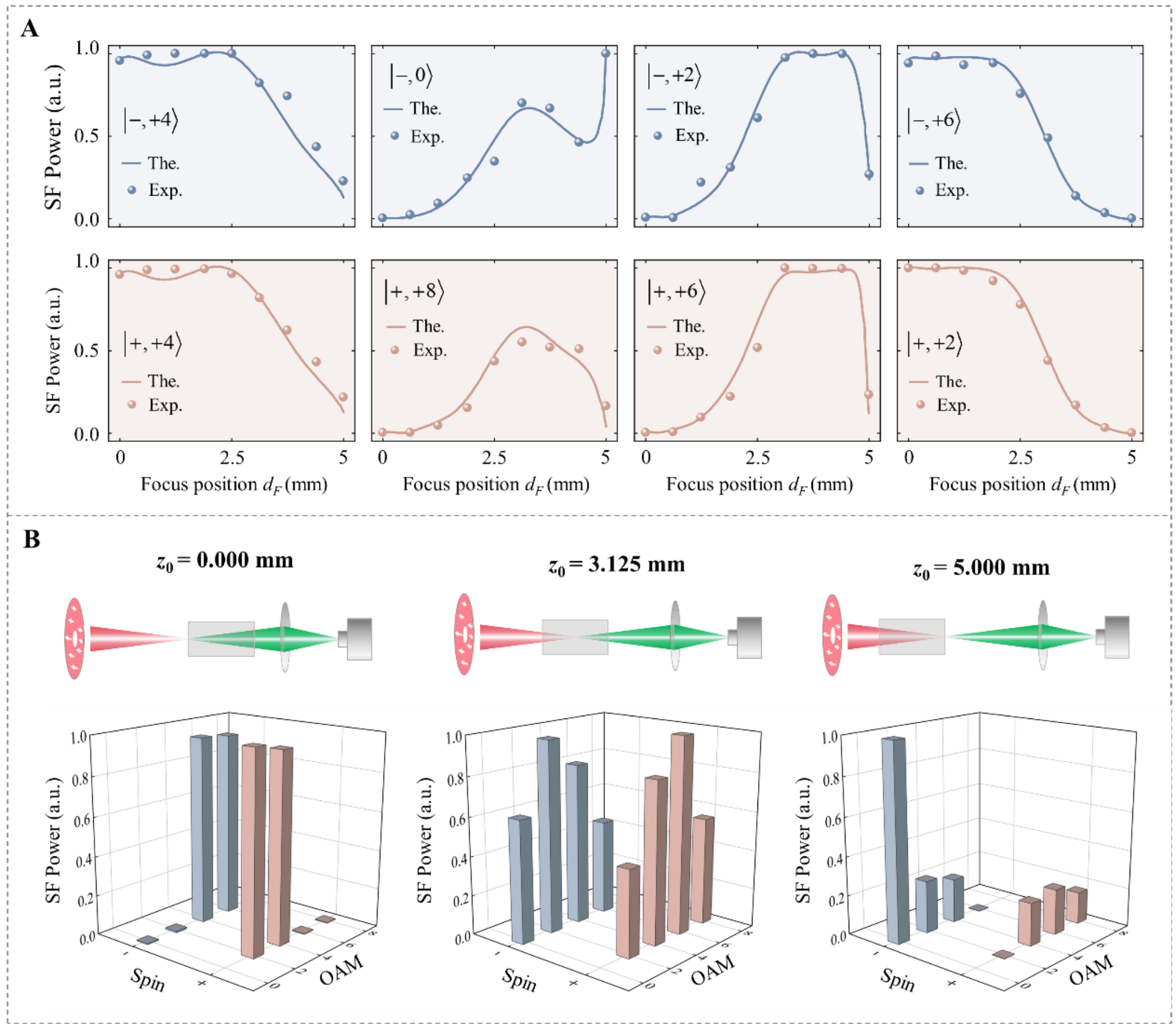

**Fig. 4. Spin-OAM spectrum of SF field for linearly polarized FF beams.** (**A**) Theoretical and experimental SHG Power evolution along with focus position $d_F$; (**B**) Spin-OAM spectra of the SF field at $d_F$=0.000 mm, 3.125 mm, and 5.000 mm.

## Optical inference platform for classification tasks

Crucially, the nonlinear wave mixing between the input FF and output SF spin-OAM states inherently instantiates a physical optical kernel function, executing high-dimensional feature expansion analogous to an SVM. In traditional computational SVMs, a mathematical kernel computes inner products in an expanded, high-dimensional space (*35-37*). In our architecture, the SHG-SOC cascade naturally synthesizes both the self- and cross-terms of the input variables. Consequently, the nonlinear crystal serves as a physical instantiation of this exact kernel function. Taking $\mathbf{E}_{in} = |+,-2\rangle + |-,+2\rangle$ as an example (Fig. 1B), an input comprising two modes, representing a 2-dimensional feature space, can be mapped onto 12 output modes, thereby expanding into a 12-dimensional feature space. Consequently, this physical mechanism can be directly harnessed to perform as an optical inference platform for classification tasks. Here we adapt the popular Iris classification task to illustrate this process (*45*). The four-dimensional input data vectors from the Iris dataset are encoded into the relative weights of the spin-OAM components (petal length encoded by $|+,-2\rangle$, petal width encoded by $|+,0\rangle$, sepal width encoded by $|-,0\rangle$, sepal length encoded by $|-,+2\rangle$). The whole experimental schema is shown as Fig. 5A (See MATERIALS AND METHODS for details). The deterministic physical system governed by nonlinear optics serves as the kernel function for feature expansion. This cascaded SOC process maps the 4-dimensional input feature space into a linear separable multi-dimensional optical OAM spectrum, and meanwhile the resulting linear decision boundary can be projected back into a low-dimensional observable space. We find that projecting onto the fundamental Gaussian modes of the orthogonal circularly polarized components perfectly reconstructs this boundary, thereby allowing each Iris class to be identified simply by detecting these fundamental modes.

The optical kernel is optimized by calibrating the physical mapping parameters. Specifically, the optical powers of the RCP and LCP fundamental modes for all 100 sets of training data are recorded at various focal planes. By calculating the cross-entropy between the output optical powers and the target classes, the focal position $d_F$=5.000 mm—which yields the minimum average cross-entropy—is selected to finalize the physical kernel configuration. Subsequently, 18 sets of test data are input to verify the feature-mapping capacity of the system. The resulting power distributions of the RCP and LCP Gaussian states ($|+,0\rangle$ and $|-,0\rangle$) are plotted in Fig. 5B. These form three distinct and well-separated clusters, successfully corresponding to the Setosa, Versicolor, and Virginica categories. More details are discussed in Supplementary Materials (Section III). This demonstrates that the nonlinear cascaded SOC process has successfully mapped the linearly inseparable input data into a high-dimensional feature space where the classes are strictly linearly separable.

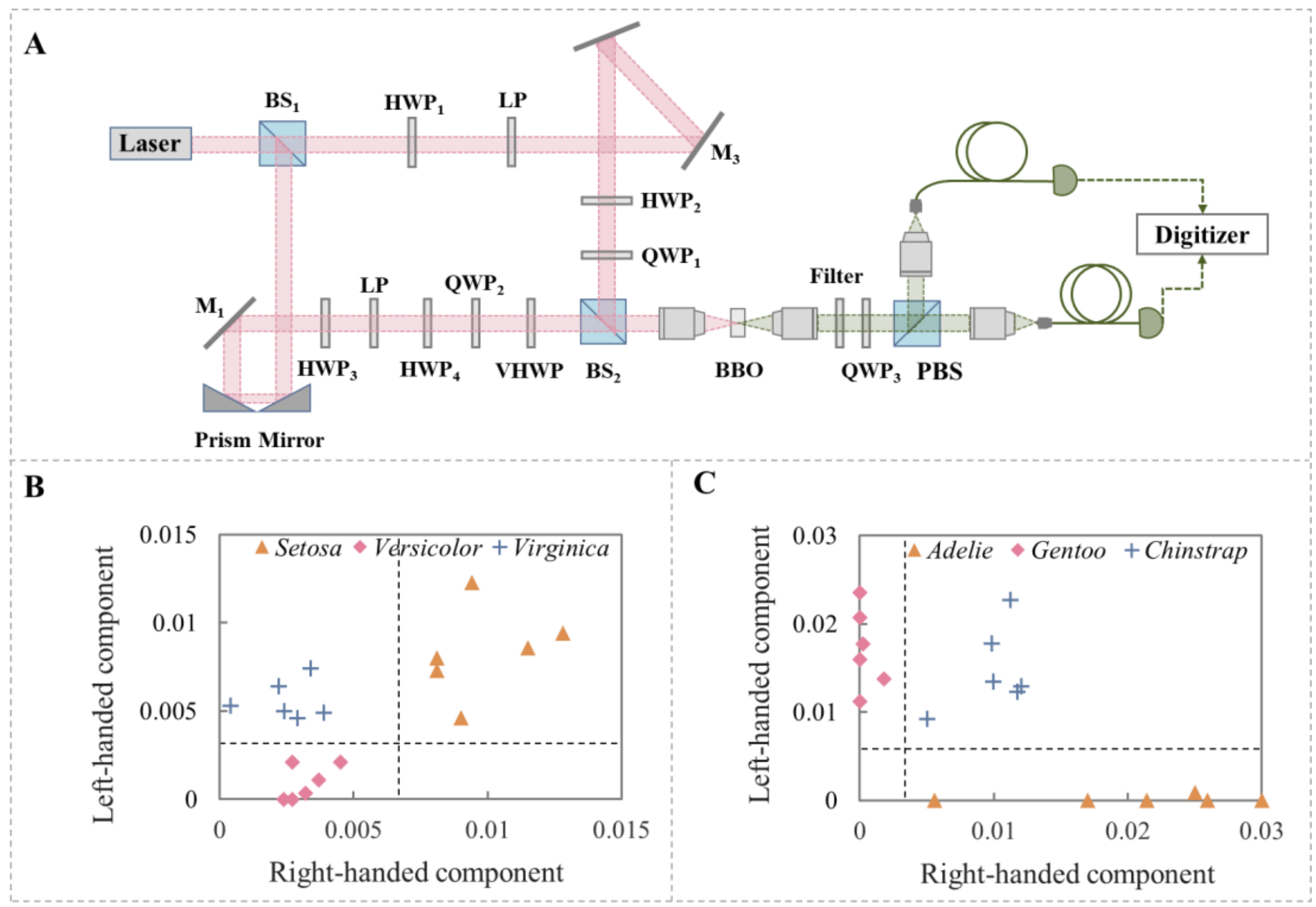

**Fig. 5. Optical inference platform for classification task.** (**A**) Experimental scheme of the photonic classification task; (**B**) Powers of right- and left-handed circular polarization component for different classes in the Iris dataset; (**C**) Right- and left-handed circular polarization component powers for different classes in the Palmer penguins dataset.

Furthermore, to highlight the reconfigurability of this architecture, we apply the all-optical inference platform to the Palmer Penguins classification task (*46*). Crucially, system reconfiguration requires no complex hardware modifications, seamlessly retaining the exact experimental setup used for the Iris task. To accommodate the penguin dataset, the input dimension is expanded to six components (island, sex, bill length, bill depth, flipper length, and body mass), with categorical features (specifically island and sex) sequence-encoded prior to optical modulation. These six-dimensional vectors are mapped onto the relative weights of the incident spin-OAM states, constituting the low-dimensional input feature space. Upon propagating through the nonlinear crystal, the optical field undergoes high-dimensional feature expansion driven by the cascaded SOC. Following the aforementioned calibration strategy, the cross-entropy is evaluated for 150 sets of training data across varying longitudinal positions to determine the optimal focal plane at $d_F$ =4.550 mm. Finally, the penguins are categorized into Adelie, Gentoo, and Chinstrap classes (Fig. 5C). More details are discussed in Supplementary Materials (Section IV). Thus, the dynamic reconfiguration of the physical mapping matrix can be efficiently realized simply by expanding the basis of the incident spin-OAM states and spatially translating the relative position of the nonlinear crystal.

## DISCUSSION

We have proposed and experimentally realized a physically reconfigurable all-optical inference architecture governed by continuously tunable cascaded SOC within a single nonlinear crystal. Under linearly polarized excitation, the sequential cascaded SOC engages synchronously with both degenerate and nondegenerate second-harmonic generation. The resulting nonlinear evolution deterministically expands the OAM spectrum, providing a vast high-dimensional orthogonal basis for feature extraction. Crucially, this mechanism intrinsically implements an optical analog of a kernel function: by simply translating the longitudinal position of the crystal, we can continuously control the physical mapping matrix, thereby projecting low-order input states into a highly tunable, linearly separable high-dimensional feature space. Based on this scheme, we successfully executed complex multivariable classification tasks, including the Iris flower and Palmer Penguins datasets.

Beyond task-specific demonstrations, this work introduces several noteworthy advances. First, unlike conventional optical computing approaches that rely on bulky spatial light modulators or fixed holographic masks, our platform achieves reconfigurability through a single, passive mechanical degree of freedom—crystal translation—enabling dynamic adaptation without requiring reprogrammable elements. Second, the intrinsic nonlinearity of the cascaded SOC process eliminates the need for separate activation functions or digital post-processing, offering a compact and low-power alternative to electronic and hybrid optoelectronic systems. Third, the physical mapping matrix can be scaled by simply expanding the basis of incident spin-OAM states, making the architecture naturally extensible to higher-dimensional tasks with minimal additional complexity.

Nevertheless, several aspects warrant further investigation. The current implementation relies on a bulk nonlinear crystal, which may limit integration density; future work could explore waveguide or metasurface-based designs for chip-scale deployment. Additionally, while the demonstrated classification accuracy is promising, the inference reconfiguration speed is primarily constrained by the response time of the translation stage rather than the optical computation itself. Integrating fast beam-steering or electro-optic positioning could enable real-time reconfigurable computing. Furthermore, extending this physics-algorithm analogy to more complex machine learning models—such as support vector regression or multi-layer perceptron—represents an intriguing direction.

In summary, our work successfully maps abstract mathematical feature expansion onto a deterministic nonlinear optical dynamical process, establishing a scalable, hardware-efficient, and low-power pathway for high-dimensional photonic computing. We anticipate that this architecture will inspire further synergies between nonlinear optics and neuromorphic photonics, paving the way for all-optical inference engines capable of adaptive, high-speed information processing.

## MATERIALS AND METHODS

### Measurement of spin-OAM spectrum of SF field

In experimental scheme, the HOP beam is prepared by a series of wave plates (half wave plate (HWP), quarter wave plate (QWP), vortex HWP (VHWP)), and then is focused by an objective (50x) (*47*). A z-cut barium metaborate (BBO) crystal is chosen as nonlinear crystal. The generated SF signal is collimated by another objective (20x), and the FF is removed by a filter. The SF fields are analyzed by a CCD with a polarization analyzer. To measure the spin-OAM spectrum of SF field, the SF field passes through a circular polarizer (containing a QWP and a linear polarizer) and a VPP, and then is focused into a single-mode fiber, which is connected with an optical power meter. By adjusting the circular polarizer and VPP, the optical powers for different spin-OAM stats can be and measured.

### Experimental scheme of classification task

In experimental scheme of classification task (Fig. 5A), a laser with wavelength 1064 nm is split into two paths in an interferometric configuration. The input spin-OAM states are meticulously prepared using a combination of linear polarizers (LPs), wave plates, and VHWP. By adjusting the angle of HWP, the multiple dimensional feature vector is converted into a specific phase pattern. The spatially modulated FF beam is then focused into the BBO crystal, where the SHG- SOC cascade establishes a physical analogue to an optical kernel function. The generated SF light is separated from the fundamental one by a filter. The RCP and LCP components of SF light are separated into two paths by a QWP and a polarization beam splitter (PBS). The fundamental modes of two circular polarization components are collected respectively into two single-mode fibers. The optical powers of the fundamental modes of RCP and LCP components for the whole training dataset are recorded at different focal planes. By calculating the cross-entropy based on the mapping between optical powers and target classes (*48-50*), the focal position can be determined with the minimum average cross-entropy to finalize the kernel function configuration.

## REFERENCES

1. L. Allen, M. W. Beijersbergen, R. J. C. Spreeuw, J. P. Woerdman, Orbital angular momentum of light and the transformation of Laguerre-Gaussian laser modes, *Phys. Rev. A* **45**, 8185 (1992).
2. A. M. Yao, M. J. Padgett, Orbital angular momentum: origins, behavior and applications, *Adv. Opt. Photonics* **3**, 161-204 (2011).
3. K. Y. Bliokh, F. J. Rodríguez-Fortuño, F. Nori, A. V. Zayats, Spin-orbit interactions of light, *Nat. Photonics* **9**, 796-808 (2015).
4. L. B. Ma, S. L. Li, V. M. Fomin, M. Hentschel, J. B. Götte, Y. Yin, M. R. Jorgensen, O. G. Schmidt, Spin-orbit coupling of light in asymmetric microcavities, *Nat. Commun.* **7**, 10983 (2016).
5. S. Gong, F. Alpeggiani, B. Sciacca, E. C. Garnett, L. Kuipers, Nanoscale chiral valley-photon interface through optical spin-orbit coupling, *Science* **359**, 443-447 (2018).
6. L. Mkhumbuza, P. Ornelas, A. Dudley, I. Nape, K. A. Forbes, Topological control of chirality and spin with structured light, *Light Sci. Appl.* **15**, 214 (2026).
7. Y. Zhao, J. S. Edgar, G. D. M. Jeffries1, D. McGloin, D. T. Chiu, Spin-to-orbital angular momentum conversion in a strongly focused optical beam, *Phys. Rev. Lett.* **99**, 073901, (2007).
8. J. Chen, L. Yu, C. Wan, Q. Zhan, Spin-orbit coupling within tightly focused circularly polarized spatiotemporal vortex wavepacket, *ACS Photonics* **9**, 793-799 (2022).
9. E. Brasselet, Y. Izdebskaya, V. Shvedov, A. S. Desyatnikov, W. Krolikowski, Y. S. Kivshar, Dynamics of optical spin-orbit coupling in uniaxial crystals, *Opt. Lett.* **34**, 1021-1023 (2009).
10. L. Marrucci, C. Manzo, D. Paparo, Optical spin-to-orbital angular momentum conversion in inhomogeneous anisotropic media, *Phys. Rev. Lett.* **96**, 163905 (2006).

11. A. Sreedharan, N. K. Viswanathan, Spin-orbit coupling mediated transverse spin mode rotation in a uniaxial crystal, *Opt. Lett.* **47**, 3768-3771 (2022).
12. S. Xiao, J. Wang, F. Liu, S. Zhang, X. Yin, J. Li, Spin-dependent optics with metasurfaces, *Nanophotonics* **6**, 215-234 (2017).
13. R. C. Devlin, A. Ambrosio, N. A. Rubin, J. P. B. Mueller, F. Capasso, Arbitrary spin-to-orbital angular momentum conversion of light, *Science* **358**, 896-901 (2017).
14. Q. Jiang, J. Laverdant, C. Symonds, A. Pham, C. Leluyer, S. Guy, A. Drezet, J. Bellessa, Metasurface for reciprocal spin-orbit coupling of light on waveguiding structures, *Phys. Rev. Appl.* **10**, 014014 (2018).
15. X. Zhou, Z. Xiao, H. Luo, S. Wen, Experimental observation of the spin Hall effect of light on a nanometal film via weak measurements, *Phys. Rev. A* **85**, 043809 (2012).
16. W. Zhu, S. Zhang, X. Liang, H. Zheng, Y. Zhong, J. Yu, Z. Chen, L. Zhang, Joint spatial weak measurement with higher-order Laguerre-Gaussian point states, *Opt. Express* **30**, 17848 (2022).
17. J. Wang, J. Yang, I. M. Fazal, N. Ahmed, Y. Yan, H. Huang, Y. Ren, Y. Yue, S. Dolinar, M. Tur, A. E. Willner, Terabit free-space data transmission employing orbital angular momentum multiplexing, *Nat. Photonics* **6**, 488-496 (2012).
18. P. Gregg, P. Kristensen, A. Rubano, S. Golowich, L. Marrucci, S. Ramachandran, Enhanced spin orbit interaction of light in highly confining optical fibers for mode division multiplexing, *Nat. Commun.* **10**, 4707 (2019).
19. O. G. Rodríguez-Herrera, D. Lara, K. Y. Bliokh, E. A. Ostrovskaya, C. Dainty, Optical nanoprobing via spin-orbit interaction of light, *Phys. Rev. Lett.* **104**, 253601 (2010).
20. G. Araneda, S. Walser, Y. Colombe, D. B. Higginbottom, J. Volz, R. Blatt, A. Rauschenbeutel, Wavelength-scale errors in optical localization due to spin-orbit coupling of light, *Nat. Phys.* **15**, 17-21 (2019).
21. J. Petersen, J. Volz, A. Rauschenbeutel, Chiral nanophotonic waveguide interface based on spin-orbit interaction of light, *Science* **346**, 67-71 (2014).
22. M. Thomaschewski, Y. Yang, C. Wolff, A. S. Roberts, S. I. Bozhevolnyi, On-chip detection of optical spin-orbit interactions in plasmonic nanocircuits, *Nano Lett.* **16**, 1166-1171 (2019).
23. S. Sukhov, V. Kajorndejnukul, R. R. Naraghi, A. Dogariu, Dynamic consequences of optical spin-orbit interaction, *Nat. Photonics* **9**, 809-812 (2015).
24. Y. Zhang, Q. Lin, Z. Zhuang, F. Lin, L. Hong, Z. Che, L. Zhuo, Y. Li, L. Zhang, D. Zhao, Dynamics of dual-orbit rotations of nanoparticles induced by spin-orbit coupling, *Nanophotonics* **14**, 833-843 (2025).
25. Y. Tang, K. Li, X. Zhang, J. Deng, G. Li, E. Brasselet, Harmonic spin-orbit angular momentum cascade in nonlinear optical crystals, *Nat. Photonics* **14**, 658-662 (2020).
26. G. Li, L. Wu, K. F. Li, S. Chen, C. Schlickriede, Z. Xu, S. Huang, W. Li, Y. J. Liu, E. Y. Pun, T. Zentgraf, K. W. Cheah, Y. Luo, S. Zhang, Nonlinear metasurface for simultaneous control of spin and orbital angular momentum in second harmonic generation, *Nano Lett.* **17**, 7974 (2017).
27. K. Nagai, T. Okamoto, Y. Shinohara, H. Sanada, Katsuya Oguri, High-harmonic spin-orbit angular momentum generation in crystalline solids preserving multiscale dynamical symmetry, *Sci. Adv.* **10**, 7315 (2024).
28. G. Li, S. Chen, N. Pholchai, B. Reineke, P. W. H. Wong, E. Y. B. Pun, K. W. Cheah, T. Zentgraf, S. Zhang, Continuous control of the nonlinearity phase for harmonic generations, *Nat. Commun.* **14**, 607-612 (2015).
29. Y. Wu, Y. Tang, Z. Hu, L. Feng, G. Guo, X. Ren, G. Li, Optical spin-orbit interaction in spontaneous parametric downconversion, *Optica* **10,** 538-543 (2023).
30. J. Shi, K. Jing, L. Li, W. Zhang, T. Zhang, X. He, Second harmonic generation of optical spin-orbit interactions in hybrid plasmonic nanocircuits, *Nanophotonics* **14**, 1003-1007 (2025).

31. D. de Ceglia, L. Coudrat, I. Roland, M. A. Vincenti, M. Scalora, R. Tanos, J. Claudon, J. Gérard, A. Degiron, G. Leo, C. De Angelis, Nonlinear spin-orbit coupling in optical thin films, *Nat. Commun.* **15**, 1625 (2024).
32. K. Konishi, T. Higuchi, J. Li, J. Larsson, S. Ishii, M. Kuwata-Gonokami, Polarization-controlled circular second-harmonic generation from metal hole arrays with threefold rotational symmetry, *Phys. Rev. Lett.* **112**, 135502 (2014).
33. Y. Tang, Z. Hu, J. Deng, K. Li, G. Li, Sequential harmonic spin-orbit angular momentum generation in nonlinear optical crystals, *Opto-Electron Adv.* **7**, 240138 (2024).
34. J. T. Pan, B. H. Zhu, L. L. Ma, W. Chen, G. Y. Zhang, J. Tang, Y. Liu, Y. Wei, C. Zhang, Z. H. Zhu, W. G. Zhu, G. Li, Y. Q. Lu, N. A. Clark, Nonlinear geometric phase coded ferroelectric nematic fluids for nonlinear soft-matter photonics, *Nat. Commun.* **15**, 8732 (2024).
35. C. Cortes, V. Vapnik, Support-vector networks, *Mach. Learn.* **20**, 273-297 (1995).
36. S. Amari, S. Wu, Improving support vector machine classifiers by modifying kernel functions, *Neural Netw.* **12**, 783-789 (1999).
37. B. Scholkopf, S. Mika, C. J. C. Burges, P. Knirsch, K. R. Muller, G. Ratsch, A. J. Smola, Input space versus feature space in kernel-based methods, *IEEE Trans. Neural Netw.* **10**, 1000-1017 (1999).
38. M. Awad, R. Khanna, Efficient learning machines: theories, concepts, and applications for engineers and system designers (Springer, Berlin, 2015).
39. D. Valkenborg, A. J. Rousseau, M. Geubbelmans, T. Burzykowski, Support vector machines, *Am. J. Orthod. Dentofacial Orthop.* **164**, 754-757 (2023).
40. A. Ciattoni, B. Crosignani, P. Di Porto, Vectorial theory of propagation in uniaxially anisotropic media, *J. Opt. Soc. Am. A* **18**, 1656 (2001).
41. A. Yariv, P. Yeh, Optical waves in crystals (Wiley, New York, 1984).
42. A. Ciattoni, G. Cincotti, C. Palma, Propagation of cylindrically symmetric fields in uniaxial crystals, *J. Opt. Soc. Am. A. Opt. Image Sci. Vis.* **19**, 792-796 (2002).
43. G. Milione, H. I. Sztul, D. A. Nolan, R. R. Alfano, Higher-order Poincaré sphere, Stokes parameters, and the angular momentum of light, *Phys. Rev. Lett.* **107**, 053601 (2011).
44. S. Chen, X. Zhou, Y. Liu, X. Ling, H. Luo, S. Wen, Generation of arbitrary cylindrical vector beams on the higher order Poincaré sphere, *Opt. Lett.* **39**, 5274-5276 (2014).
45. R. A. Fisher, The use of multiple measurements in taxonomic problems, *Ann. Eugen.* **7**, 179-188 (1936).
46. K. B. Gorman, T. D. Williams, W. R. Fraser, Ecological sexual dimorphism and environmental variability within a community of Antarctic penguins (genus Pygoscelis), *PLoS One* **9**, e90081 (2014).
47. D. Naidoo, F. S. Roux, A. Dudley, I. Litvin, B. Piccirillo, L. Marrucci, A. Forbes, Controlled generation of higher-order Poincaré sphere beams from a laser, *Nat. Photonics* **10**, 327-332 (2016).
48. X. Lin, Y. Rivenson, N. T. Yardimci, M. Veli, Y. Luo, M. Jarrahi, A. Ozcan, All-optical machine learning using diffractive deep neural networks, *Science* **361**, 1004-1008 (2018).
49. T. W. Hughes, M. Minkov, Y. Shi, S. Fan, Training of photonic neural networks through in situ backpropagation and gradient measurement, *Optica* **5**, 864-871 (2018).
50. L. G. Wright, T. Onodera, M. M. Stein, T. Wang, D. T. Schachter, Z. Hu, P. L. McMahon, Deep physical neural networks trained with backpropagation, *Nature* **601**, 549-555 (2022).